\documentclass[aps,prl,twocolumn,showpacs,10pt,floatfix,superscriptaddress,floatfix, longbibliography]{revtex4-1}
\usepackage{epsf}
\usepackage{bm}
\usepackage{hyperref}
\usepackage{amsfonts}
\usepackage{amssymb}
\usepackage{amsmath,mathtools}
\usepackage{array}
\usepackage{enumerate,dsfont}
\usepackage{dcolumn,multirow}
\usepackage[utf8]{inputenc}
\usepackage{latexsym}
\usepackage{xcolor}
\usepackage{tikz-cd}
\usepackage{ulem}
\renewcommand{\emph}[1]{{\it #1}}
\renewcommand{\em}{\it}
\newcommand{\overbar}[1]{2 #1}
\newcommand{\aH}{H_\mathrm{aux}}

\newcommand{\cmark}{\checkmark}%
\newcommand{\xmark}{0}%
\newcommand{\W}[1]{{\text{Wi}_{#1}}}
\newcommand{\Ch}[1]{{\text{Ch}_{#1}}}
\newcommand{\vk}{\vec k}

\newcommand{\vn}{\vec n}

\renewcommand{\vr}{\vec r}

\newcommand{\ZZ}{\mathbb{Z}}

\newcommand{\ket}[1]{{| #1 \rangle}}

\usepackage{lipsum}

\renewcommand{\vec}[1]{\mathbf{#1}}

\begin{document}

\title{Topology of ultra-localized insulators and superconductors}

\author{Bastien Lapierre}
\affiliation{Department of Physics, Princeton University, Princeton, New Jersey, 08544, USA}
\author{Luka Trifunovic}
\affiliation{Laboratoire de Physique Théorique, CNRS and Université de Toulouse, 31400 Toulouse, France}
\author{Titus Neupert}
\affiliation{Department of Physics, University of Zurich, Winterthurerstrasse 190, CH-8057 Zürich, Switzerland}
\author{Piet W. Brouwer}
\affiliation{Dahlem Center for Complex Quantum Systems and Physics Department, Freie Universit\"at Berlin, Arnimallee 14, 14195 Berlin, Germany}

\date{\today}

\begin{abstract}
The topology of an insulator can be defined even when \emph{all} eigenstates of
the system are localized --- an extreme case of Anderson insulators that we
call ultra-localized. We derive the classification of such ultra-localized
insulators in all symmetry classes and dimensions. We clarify their
bulk-boundary correspondence and show that ultra-localized systems are in many
instances phases of matter not described by the known classification of
topological insulators and superconductors. As a consequence,
we clarify which conventional topological phases are Wannierizable, and which
topological phases cannot exist without delocalized states.
\end{abstract}
\maketitle

\textit{Introduction} --- Anderson insulators and band insulators share the key physical property that they do not conduct electrical current. The mechanisms by which electrons are inhibited from propagating are, however, fundamentally different. In Anderson insulators disorder localizes electron orbits, and in band insulators the ionic potential creates energetically ``forbidden'' intervals for electron waves, the band gaps. 

Band insulators are thus only insulating at very specific electron filling, while Anderson insulators can be insulating over entire ranges of fillings -- depending on the disorder strength, symmetry and dimension of the system~\cite{RevModPhys.80.1355}. At very strong disorder, Anderson insulators are \textit{insulating over their whole spectrum}. We will refer to such fully insulating systems as ``ultra-localized insulators'', or ``ultra-insulators'' (UI) for short. While in one-dimension all Anderson insulators are ultra-insulators, in higher dimensions this requires a large but finite disorder strength. Ultra-insulators are non-conducting even at arbitrary temperature or in non-equilibrium settings such as under strong applied voltages. Thus, the notion of ultra-insulator is naturally relevant to 
periodically driven systems~\cite{PhysRevX.3.031005,PhysRevX.6.021013, PhysRevLett.119.186801}, where stabilizing insulating phases is often a challenging task.

\begin{figure}
\includegraphics[width=0.48\textwidth]{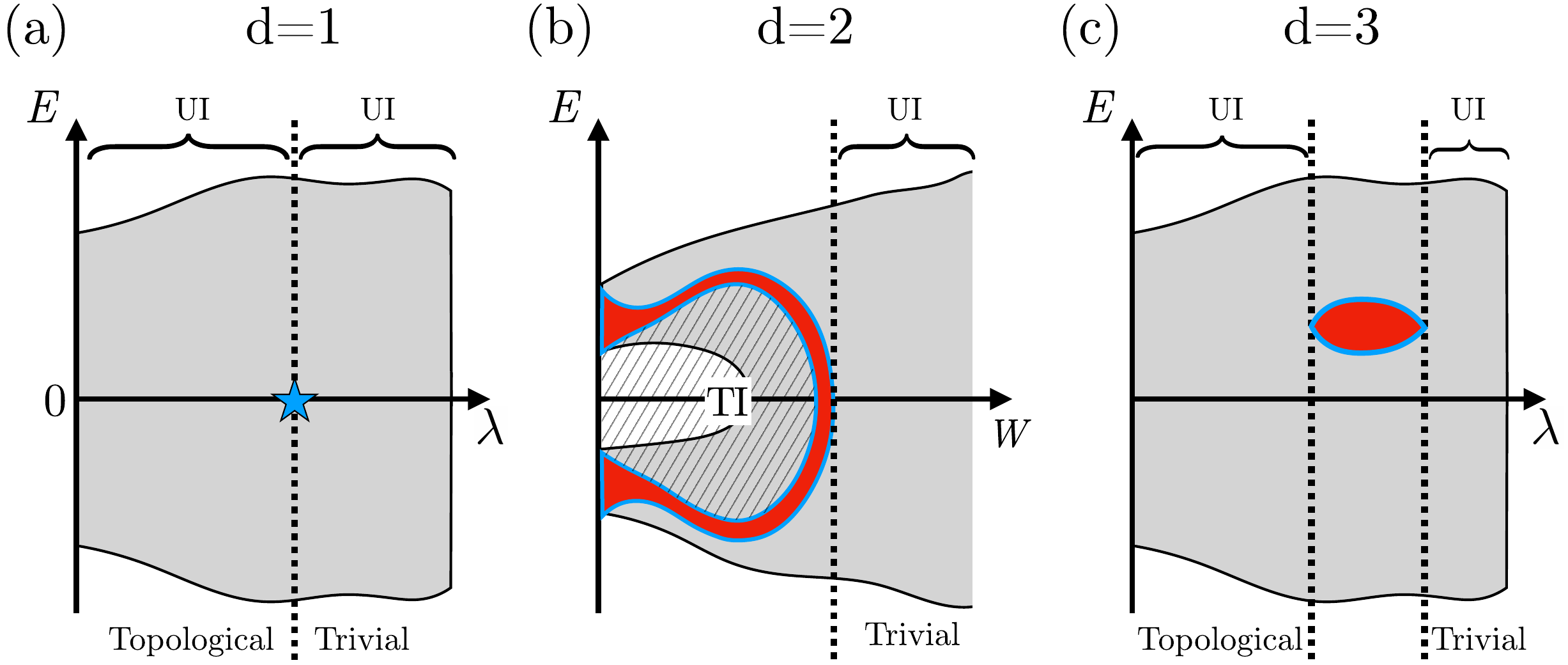}
\caption{Localization properties of eigenstates of generic disordered systems across topological phase transitions. (a) Phase transition between a topological (ultra localized) insulator and a trivial insulator in $d=1$, driven by a parameter $\lambda$. In both phases all states are localized (grey), and critical states (blue) appear at zero energy \cite{Essin_2015,bagrets2016} at the transition point. (b) Phase transition out of a disordered TI (hatched region) in $d=2$, not deformable into an ultra-insulator. The states in red are metallic. A trivial ultra-insulator emerges at strong disorder as the delocalized states pair-annihilate, while TIs appear only at specific fillings and require bulk delocalized states. Here, $W$ is the disorder strength. (c) Transition between a topological and a trivial ultra-insulator in $d=3$. There is an extended critical region characterized by delocalized states appearing somewhere in the bulk spectrum.
}
\label{Fig:ultra_ins1}
\end{figure}

A major development in the theory of band insulators is the classification of non-interacting topological insulators (TIs), insulating phases that feature boundary signatures such as backscattering-free edge modes that are robust to disorder \cite{hasan2010, RevModPhys.83.1057, Bernevig+2013}.  
Disregarding crystalline symmetries, insulators and superconductors have been topologically classified according to the tenfold-way, which provides a ``periodic table of topological phases of matter''~\cite{kitaev2009, ryu2010}.
This topological classification of band insulators further applies to Anderson insulators, where robust delocalized bulk states stabilize a nontrivial phase called topological Anderson insulator or equivalently disordered TI~\cite{PhysRevLett.102.136806, PhysRevLett.103.196805, prodan2011, PhysRevB.80.165316}, see Fig.~\ref{Fig:ultra_ins1}(b). 
On the other hand, ultra-insulators are often considered as constituting a single, topologically trivial phase of matter, as all their bulk states are localized.

It is worth asking whether ultra-insulators can be topological, too. By their definition, the notion of topological equivalence between ultra-insulators has to be a \textit{global} property of the spectrum, in stark contrast to the usual notion of topology in band or Anderson insulators that makes reference to a gap or mobility gap at a Fermi level. A topological transition between ultra-insulators requires delocalized states to appear \textit{anywhere} in the bulk spectrum, see Fig.~\ref{Fig:ultra_ins1}(c), rather than at a Fermi energy.

With this understanding, we can immediately conclude that topological ultra-insulators exist in $d=1$. There, all conventional topological phases are known to be localizable~\cite{kohn1959}, i.e., they can be continuously  deformed into ultra-insulators without closing the mobility gap at the Fermi level. Transitions between topological phases in $d=1$
necessitate critical (extended) states~\cite{bagrets2016}, see Fig.~\ref{Fig:ultra_ins1}(a). In contrast, in $d=2$ all known  topological insulators --- such as the quantum (spin) Hall effect --- crucially require the existence of bulk critical (delocalized) states~\cite{chalker1988, RevModPhys.67.357, onoda2007, morimoto2015, PhysRevB.79.045321, PhysRevLett.98.076802}, as illustrated on Fig.~\ref{Fig:ultra_ins1}(b), leaving no room for topological ultra-insulators among the known topological phases. In $d=3$, the connection between localizability and tenfold-way topology has not been made. It  therefore remains elusive to what extent topological ultra-insulators  exist above $d=1$. They would realize distinctly characterized strong, fully localized topological phases.

In this letter, we comprehensively address this question by constructing the classification of topological ultra-insulators and ultra-localized superconductors (TUIs) in any symmetry class and dimension, motivated by the recent finding of a TUI in $d=3$~\cite{lapierre2022}\footnote{We note that the notion of topologically localized insulator introduced in~\cite{lapierre2022} coincides with the notion of TUI studied in the present work.}. 
This classification exhibits all topologically distinct ultra-insulators. 
We show that, unlike conventional TIs [see Fig.~\ref{Fig:ultra_ins1}(b)], the anomalous boundaries of three dimensional TUIs host delocalized states that survive the addition of arbitrarily large boundary disorder.
As a consequence of our classification, we identify all conventional topological insulators and superconductors that can be ultra-localized.

\textit{Phenomenology of TUIs} --- 
Similarly to conventional TIs, where the condition of being an insulator is broken on their boundaries, the boundaries of TUIs violate the ultra-insulator condition. In other words, the boundary of a TUI hosts delocalized states which may occur at one energy or in an energy window that depends on details of the surface termination.

\begin{figure}[t]
\includegraphics[width=0.3\textwidth]{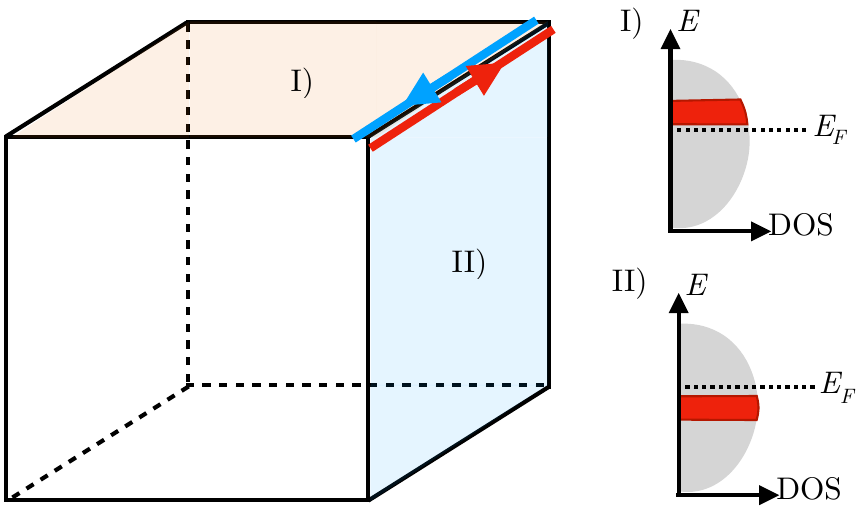}
 \caption{Sketch of the surface of a time-reversal invariant TUI in $d=3$. While its bulk consists only of fully localized orbitals (grey states), its anomalous surface can be understood as a single band of a $\mathbb{Z}_2$ TI in $d=2$.
 In particular, if the energy windows (red) supporting delocalized states for two adjacent surfaces I and II do not overlap, delocalized helical hinge states between the two surfaces are found at any energy $E_{\mathrm{F}}$ between the red windows of delocalized surface states.
 }
\label{Fig:ultra_ins2}
\end{figure}

As we show below, in the presence of time-reversal symmetry $T^2=-1$, TUIs possess a $\mathbb{Z}_2$ index in $d=3$. The states on the two-dimensional surfaces of such TUI 
resemble one ``band'' of a two-dimensional 
$\mathbb{Z}_2$ topological insulator: there is an extended energy window of delocalized states that carry a nontrivial $\ZZ_2$ invariant, as illustrated on Fig.~\ref{Fig:ultra_ins2}. For true two-dimensional $\ZZ_2$-insulators, these delocalized states can be localized by increasing bulk disorder, as the states carrying opposite spin-Chern number eventually pair-annihilate~\cite{prodan2011}, leading to a trivial ultra-insulator, see Fig.~\ref{Fig:ultra_ins1}b. On the other hand, the anomalous surfaces of a TUI persist turning into an UI for arbitrarily large surface disorder, as there is a single band of a $\ZZ_2$ insulator per surface, and different surfaces are separated by a fully localized bulk. Importantly, this implies that the surface topology of TUIs also manifests when the system is fully filled, contrary to conventional topological phases.

The energy window in which delocalized states are supported on the surface is non-universal, and thus generically differs between surfaces.
This naturally leads to helical hinge states at intermediate energies, as second-order topological modes of TUIs (see Fig~\ref{Fig:ultra_ins2}).
While similar hinge states may also appear on the surface of second order extrinsic TIs, their topological responses can be shown to be distinct when probed in non-equilibrium settings~\cite{lapierre2022}. As we will later see, there also exist TUI systems whose boundary cannot be related to a (non-localizable, see later below) TI in the same symmetry class and one dimension lower.

\textit{Classification} --- We consider a $d$-dimensional TUI on a lattice $\mathcal{L}$ with sites $\vr\in\mathcal{L}$ in $d$-dimensional Euclidean space, assuming one orbital per site for simplicity. By definition, the system is characterized by a localized bulk spectrum. It follows that its eigenstates $|\psi_{\vn} \rangle$ can be indexed by  $d$-dimensional vectors $\vn\in \mathcal{L}$,
that can always be chosen such that the eigenstate $|\psi_{\vn }\rangle$ (at energy $E_{\vn}$) is localized near site $\vr=\vn$, although such an assignment is not unique. The eigenstates $|\psi_{\vn }\rangle$ can be seen as the columns of a unitary matrix $U_{\vr ,\vn }$,
\begin{equation}
  |\psi_{\vn}\rangle = \sum_{\vr }
  U_{\vr ,\vn } |\vr\rangle, 
\end{equation}
with $U_{\vr ,\vn } = \langle \vr  | \psi_{\vn}\rangle$ and $| \vr  \rangle$ being eigenstates of the position operator, or, in other words, of a trivial reference TUI that we call atomic limit.
It follows that one may diagonalize the Hamiltonian $H$ as
\begin{equation}
\label{eq:TUIdefin}
  H = U \Lambda U^{\dagger},
\end{equation}
where $\Lambda$ is a diagonal matrix containing the energy eigenvalues $E_{\vn}$.
Because the eigenstates of $H$ are localized, $U_{\vn ,\vr }$ decays exponentially for large $|\vr - \vn|$, {\em i.e.}, it is a {\em local} unitary matrix.

With the help of the unitary matrix $U$, one may construct the auxiliary chiral Hamiltonian~\cite{Roy_2017}
\begin{equation}
  \aH = \begin{pmatrix} 0 & U \\ U^{\dagger} & 0 \end{pmatrix}.
  \label{eq:tildeH}
\end{equation}
The Hamiltonian $\aH$ is not the physical Hamiltonian for the TUI but will be a computational tool in our derivation. 
Any symmetry and antisymmetry conditions imposed on $H$ result in symmetry conditions for the unitary $U$ and, hence, for the auxiliary Hamiltonian $\aH$. Because of the $2 \times 2$ matrix structure in Eq.\ (\ref{eq:tildeH}), $\aH$ satisfies
an additional chiral symmetry, which should be interpreted as a ``sublattice symmetry'', with $\aH$ being defined on a bipartite lattice $\mathcal{L}\cup \mathcal{L}$ consisting of the original lattice (indexed by $\vr$) and the lattice of eigenstate site labels $\vn$.
We note that $\aH$ is (i) a local Hamiltonian, because $U$ is a local unitary, (ii) it is gapped, because $U$ is a unitary matrix, and (iii) it is thus classified by the usual tenfold-way classification~\cite{ryu2010}.

We use the mapping to the auxiliary Hamiltonian \eqref{eq:tildeH} to classify TUIs in all dimensions and classes.
The complete classification is summarized in Table~\ref{tab:classification_TUIs}, and details of its derivation for each class are discussed in~\cite{SM}. 
Crucially, while the topology of TUIs is invariant under a local relabelling of the eigenstates (i.e., a relabelling that leaves the unitary $U$ local), this may change the topology of the auxiliary Hamiltonian~\eqref{eq:tildeH} in $d=1$, but this relabelling freedom is 
inconsequential in higher dimensions~\cite{SM}.
As a result, the period-eight Bott periodicity of the TUI classification sets in for $d \ge 2$ only, whereas
$d=1$ is an exceptional case that never repeats in higher dimensions.
The Bott clock, which relates topological phases of different classes and
different dimensions, is absent in our classification. Instead,
Table~\ref{tab:classification_TUIs} has a threefold structure, where the
resulting classification is associated to one of the three chiral classes AIII,
BDI, and CII. We note that for $d>1$ our classification table conicides with
the topological classification of a totally different systems, i.e., that of
point-gap non-Hermitian systems~\cite{gong2018}.

\begin{table}
\begin{center}
\resizebox{0.48\textwidth}{!}{
	\begin{tabular}[b]{l@{\extracolsep{\fill}} |c|cccccccc} \hline\hline
                                class  & $\,d=1$ & $d=2$ & $d=3$ & $d=4$ & $d=5$ & $d=6$ & $d=7$ & $d=8$ & $d=9$\\ \hline
                                AIII   & $\ZZ$ & $0$  & $\ZZ\times\ZZ$ & $0$ & $\ZZ\times\ZZ$ & $0$ & $\ZZ\times\ZZ$ & $0$ & $\ZZ\times\ZZ$\\
                                A      & $0$ & $0$ & $\ZZ$ & $0$ & $\ZZ$ & $0$ & $\ZZ$ & $0$ & $\ZZ$\\
                                DIII   & $\ZZ_2$ & $0$ & $\ZZ$  & $0$ & $\ZZ$ & $0$ & $\ZZ$ & $0$ & $\ZZ$\\
                                CI   & $0$ & $0$ & $\ZZ$ & $0$ & $\ZZ$ & $0$ &  $\ZZ$ & $0$ & $\ZZ$\\
                                \hline
                                BDI    & $\ZZ$  & $0$ & $0$ & $0$ & $2\ZZ\times2\ZZ$ & $0$ & $\ZZ_2\times\ZZ_2$ & $\ZZ_2\times\ZZ_2$ & $\ZZ\times\ZZ$\\
                                AI     & $0$ & $0$ & $0$ &$0$ & $2\ZZ$ & $0$ & $\ZZ_2$ & $\ZZ_2$ & $\ZZ$\\
                                D      & $\ZZ_2$ &  $0$ & $0$ & $0$ & $2\ZZ$ & $0$ & $\ZZ_2$ & $\ZZ_2$ & $\ZZ$\\
                                \hline
                                CII    & $2\ZZ$ & $0$ & $\ZZ_2\times\ZZ_2$ & $\ZZ_2\times \ZZ_2$ & $\ZZ\times\ZZ$ & $0$ & $0$ & $0$ & $2\ZZ\times2\ZZ$\\
                                AII    & $0$ & $0$ & $\ZZ_2$ & $\ZZ_2$ & $\ZZ$ & $0$ & $0$ & $0$ & $2\ZZ$\\
                                C   & $0$ & $0$ & $\ZZ_2$ & $\ZZ_2$ & $\ZZ$ & $0$ & $0$ & $0$ & $2\ZZ$\\
                                \hline\hline
                        \end{tabular}
}
			\caption{Classification of topological ultra-insulators and superconductors. The classification has a threefold structure (with an exception in $d=1$), with three groups associated to the three chiral classes. The bulk topological invariants associated to each of these systems are summarized in~\cite{SM}. “$2\ZZ$” indicates that the invariant takes even values only.
   \label{tab:classification_TUIs}
   }
	\end{center}
\end{table}

We now illustrate our classification scheme in class AII in $d=3$. Consider an ultra-insulator
subject to time-reversal symmetry $T^2=-1$,
\begin{equation}
H = \sigma_2 H^*\sigma_2.
\label{eq:timereversal}
\end{equation}
Eigenvalues of $H$ come in twofold (Kramers) degenerate pairs.
The same symmetry condition applies to the eigenvector matrix $U$, which makes it a symplectic matrix, and to the auxiliary chiral Hamiltonian $\aH$, which places it  in tenfold-way class CII.
Since class CII admits a $\ZZ_2$ classification in $d=3$, this carries over to TUIs in class AII.
The strategy similarly applies in classes A and AI, where the auxiliary Hamiltonians lie in classes AIII and BDI respectively. This leads to a $\mathbb{Z}$ classification for class A~\cite{lapierre2022}, while TUIs are always trivial in class AI in $d=3$.
We stress that while the existence of the  $\mathbb{Z}_2$ TUI can be anticipated from the existence of a $\mathbb{Z}_2$ TI in $d=2$, this is not always the case, e.g., topological ultra-localized superconductor in class C and $d=3$ have a $\mathbb{Z}_2$ classification, whereas conventional class-C topological superconductor in $d=2$ have a $\mathbb{Z}$ classification.

\textit{Wannier localizability of TIs} ---
Our classification of TUIs can be used to determine an important property of conventional topological insulator phases.
Certain topological phases, such as quantum (spin) Hall insulators and three-dimensional TIs, have an obstruction to Wannier localization of their bulk spectrum~\cite{chalker1988, onoda2007, morimoto2015}, and thus cannot be deformed to TUIs without closing the mobility gap. Others, such as the one-dimensional topological phases, can be adiabatically connected to an ultra-localized limit~\cite{kohn1959}.
It is then natural to ask which topological phases of the conventional TI classification can exist as ultra-localized insulators and superconductors~\cite{monaco2016,panati2018,marcelli2023}.
Since for $d > 0$ topological equivalence as TUI phases implies topological equivalence according to the TI classification, it is possible to assign a TI invariant to each TUI. The result of such an assignment is shown in Table~\ref{tab:TF}: A ``$\cmark$'' in Tab.~\ref{tab:TF} indicates that all topological phases in the corresponding symmetry class exist as TUI ({\em i.e.}, that the TI invariants calculated for TUIs span the full set of allowed TI invariants for that symmetry class), whereas a ``$\xmark$'' in Tab.~\ref{tab:TF} indicates that TUIs have trivial TI invariant.

\begin{table}
\begin{center}
\resizebox{0.48\textwidth}{!}{
	\begin{tabular}[b]{l@{\extracolsep{\fill}} |ccccccccc} \hline\hline
                                class  & $\,d=1$ & $d=2$ & $d=3$ & $d=4$ & $d=5$ & $d=6$ & $d=7$ & $d=8$ & $d=9$\\ \hline
                                A      & $0$ & $\xmark$ & $0$ & $\xmark$ & $0$ & $\xmark$ & $0$ & $\xmark$ & $0$\\
                                AIII   & $\cmark$ & $0$  & $\cmark$ & $0$ & $\cmark$ & $0$ & $\cmark$ & $0$ & $\cmark$\\
                                \hline
                                AI     & $0$ & $0$ & $0$ &$\xmark$ & $0$ & $\xmark$ & $\xmark$ & $\xmark$ & $0$\\
                                BDI    & $\cmark$  & $0$ & $0$ & $0$ & $\cmark$ & $0$ & $\cmark$ & $\cmark$ & $\cmark$\\
                                D      & $\cmark$ &  $\xmark$ & $0$ & $0$ & $0$ & $\xmark$ & $0$ & $\cmark$ & $\cmark$\\
                                DIII   & $\cmark$ & $\xmark$ & $\xmark/\cmark$  & $0$ & $0$ & $0$ & $\cmark$ & $0$ & $\cmark$\\
                                AII    & $0$ & $\xmark$ & $\xmark$ & $\xmark$ & $0$ & $0$ & $0$ & $\xmark$ & $0$\\
                                CII    & $\cmark$ & $0$ & $\cmark$ & $\cmark$ & $\cmark$ & $0$ & $0$ & $0$ & $\cmark$\\
                                C   & $0$ & $\xmark$ & $0$ & $\cmark$ & $\cmark$ & $\xmark$ & $0$ & $0$ & $0$\\
                                CI   & $0$ & $0$ & $\cmark$ & $0$ & $\cmark$ & $\xmark$ &  $\xmark/\cmark$ & $0$ & $0$\\
                                \hline\hline
                        \end{tabular}}

			\caption{Wannier localizability of conventional tenfold-way topological phases (indicated by $\cmark$). Wigner-Dyson classes are never Wannier localizable while chiral classes are always Wannier localizable. The Wannier localizability of superconducting classes depends on the precise class and dimension. The entry $\xmark/\cmark$ indicates that the phase is Wannier localizable for an even invariant but non-localizable for an odd invariant.
			\label{tab:TF}}
	\end{center}
\end{table}

For the Wigner-Dyson classes A, AI, AII, which do not have a ``special'' energy
$E = 0$, the energy eigenvalues of a TUI may be chosen such that all
eigenstates have negative energy. The resulting conventional invariants are
always trivial, as indicated in Tab.~\ref{tab:TF}.  The opposite is true for
chiral classes AIII, BDI and CII, where \textit{all} conventional topological
phases can be localized. As a consequence, the topology of, e.g., chiral TIs in
$d=3$ does not imply the existence of delocalized bulk
states~\cite{PhysRevX.14.011057}.
Whether the superconducting classes C, CI, D, and DIII are Wannier localizable
or not depends on the dimension and the symmetry class, as summarized in
Table~\ref{tab:TF} (see~\cite{SM} for details). It has been shown
recently~\cite{PhysRevX.14.011057}, that Wannier localizability implies
detachability of the boundary modes from the bulk spectrum. Comparing the
present results with the recent study that classifies detachable boundary
modes~\cite{nakamura2025}, we establish the equivalence of Wannier
localizability and detachability of the boundary modes in conventional
tenfold-way topological phases.

While topological equivalence as TUIs implies topological equivalence as conventional TI if $d > 0$, the opposite is not true. In this sense, the TUI classification is more ``refined'' than the TI classification. A nontrivial example is the $\ZZ\times\ZZ$ classification of chiral TUIs with broken time-reversal symmetry for $d=3$ (see Table~\ref{tab:classification_TUIs}), which has to be contrasted with the $\ZZ$ classification of TIs: For each (integer-valued) conventional topological invariant, an additional integer-valued topological invariant is required to characterize the TUI. This additional invariant counts the number of finite-energy critical states appearing on the boundary~\cite{SM,PhysRevX.14.011057}. Table~\ref{tab:TUIinv} of~\cite{SM} gives an overview of this ``beyond-tenfold-way'' aspect of the TUI classification for all symmetry classes and all dimensions.

\begin{figure}
\includegraphics[width=0.48\textwidth]{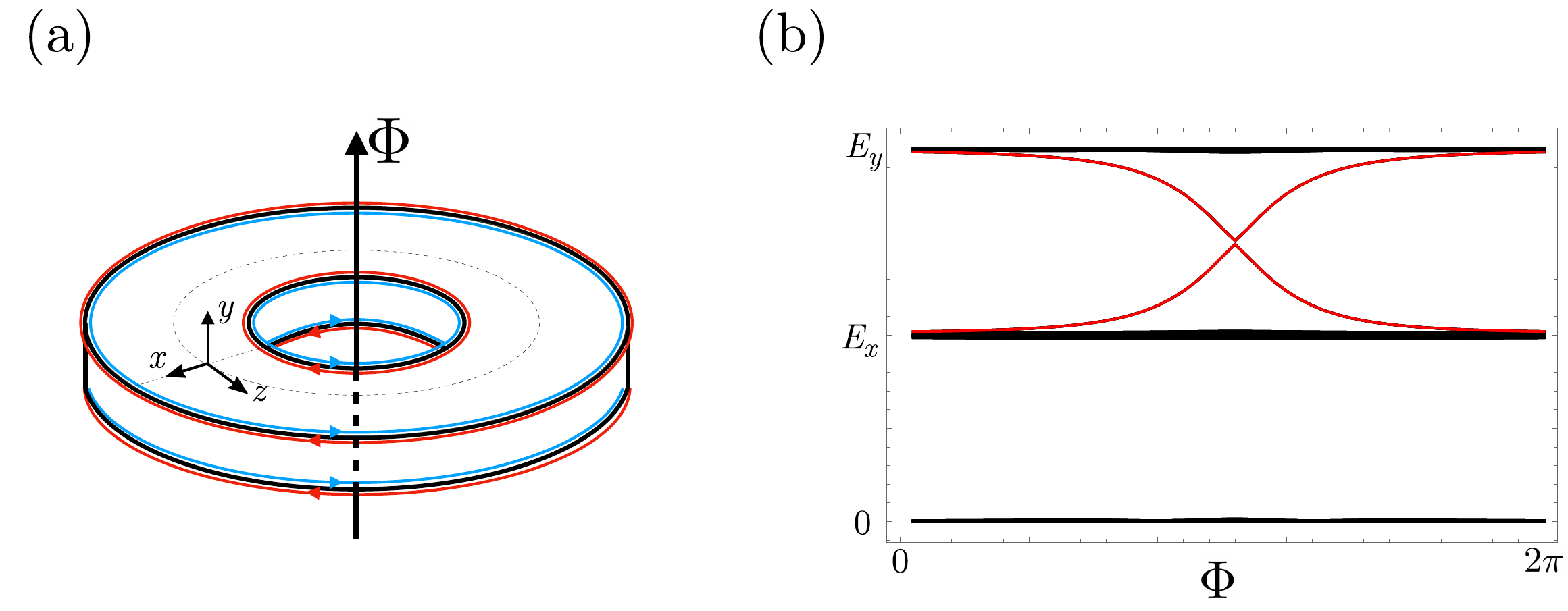}
\caption{
A TUI in a thick Corbino disk geometry, threaded by a magnetic flux $\Phi$ (a) and its spectrum as a function of $\Phi$ (b). Hinge modes are indicated in red. To bring about the hinge modes, energies of the states localized near the vertical ($\perp x$) and horizontal ($\perp y$) surfaces are shifted by $E_x$ and $E_y$, respectively. Because there are four hinges, each dispersing branch in (b) is fourfold degenerate.
}
\label{Fig:figure_hingestates}
\end{figure}

\textit{Concrete model of a TUI} ---  The classification procedure outlined above can also be used to construct concrete models of TUIs. We illustrate this for TUIs in class AII and $d=3$.
We reverse-engineer the TUI from its auxiliary Hamiltonian $\aH$, which is in tenfold-way class CII. To construct $\aH$, we start from two time-reversed copies of translation-invariant TIs in class AIII, (first line in Eq.\ (\ref{eq:HCII})), coupled with a term that respects time-reversal and chiral symmetries (second line)~\cite{liu2023elementary},
\begin{align}
  H_{\text{CII}}(\vk) =&\, 
  \sum_{i \in \{x,y,z\}} [w \Gamma_0 + v \cos (k_i) \Gamma_0 + v \sin(k_i) \Gamma_i] 
  \nonumber \\  
  &\, \mbox{}
  + \lambda s_y \tau_x [(1+\cos (k_x)) \sigma_x + \sin (k_x) \sigma_y]
.
  \label{eq:HCII}
\end{align}
Here $\Gamma_0 = s_z \tau_x \sigma_x$, $\Gamma_x = s_z \tau_x \sigma_y$, $\Gamma_y = s_z \tau_x \sigma_z$, $\Gamma_z = s_0 \tau_y \sigma_0$, where the $s_i$, $\tau_i$, and $\sigma_i$ are Pauli matrices and time-reversal $T^2=-1$ and chiral symmetry $C$ read $T =i s_y \sigma_z K$, $C = \tau_z$, respectively, $K$ being complex conjugation.
For $\lambda = 0$, this model is in a topological phase with winding number $1$ for $|v| < |w| < 3|v|$. In what follows, we set $v=2$, $w=4$, $\lambda=0.3$.
We flatten the spectrum to obtain the auxiliary Hamiltonian $\aH({\bf k})$ that has the form \eqref{eq:tildeH}, whereby the matrix $U({\bf k})$ is symplectic because of the time-reversal constraint and unitary because of the flattened spectrum. Fourier transforming to position space results in a matrix $U_{\vr,\vn}$ that is local, while preserving time-reversal symmetry and unitarity. The class-AII TUI Hamiltonian $H$ can then be extracted from $U$
using Eq.~\eqref{eq:TUIdefin}, where the eigenvalues in $\Lambda$ come in twofold degenerate pairs, which we select randomly.
As discussed previously, such TUI is characterized by helical hinge modes connecting anomalous surfaces whose spectrum consist of a single band of a two dimensional $\mathbb{Z}_2$ topological insulator.
In order to numerically exhibit such helical hinge modes, we put the system on a thick Corbino disk geometry, with periodic boundary conditions with flux $\Phi$ in the $z$ direction and open boundary conditions in the $x$ and $y$ directions, see Fig.~\ref{Fig:figure_hingestates}(a). We shift the energies of states within a localization length of the surfaces $\perp x$ and $\perp y$ by $E_x$ and $E_y$, respectively. 
Helical hinge states in the TUI then lead to Kramers degenerate hinge modes around the four hinges that continuously connect the surface bands at energies $E_x$ and $E_y$, see Fig.~\ref{Fig:figure_hingestates}(b).

\textit{Conclusions} --- 
Ultra-localized insulators, insulators that have exponentially localized states at all energies, can nevertheless be topological.
The resulting topology is a global feature of the fully localized spectrum, and only transitions between topologically inequivalent ultra-insulators lead to metallic states somewhere in the spectrum.
While these systems coincide with conventional topological matter in one
dimension, and cannot exist in two dimensions, we uncovered many new instances
of TUIs in three dimensions in different symmetry classes. As a hallmark their
topology, TUIs host anomalous delocalized surface states that resist arbitrary
surface disorder. Furthermore, TUIs provide new insights into TIs: by finding
the `intersection' of the TUI and TI classifications, we uncover many
Wannierizable TIs.

The fact that TUI and TI classifications differ is a consequence of the different objectives -- studying equivalence classes of fully localized Hamiltonians versus those of ground states of gapped Hamiltonians. 
This change of viewpoint may be fruitful when applied to time-dependent quantum systems, such as Floquet and random unitary circuits, where local unitary operators (instead of Hamiltonians) play a central role~\cite{PhysRevX.6.021013, Roy_2017, PhysRevLett.126.106805, liu2023classification, fidkowski2023pumping}. 
It would then be desirable to extend our classification scheme of local unitaries~\cite{zhang2021b} beyond the single particle description. 

The construction of TUIs presented in this work corresponds to an
infinite-disorder limit of a particular correlated disorder, instead of
commonly considered independently distributed disorder. It is an open question
whether some TUIs can be reached by adding \textit{generic} (uncorrelated)
disorder in a clean topological insulator. Nevertheless, the toy models
presented in this work are potentially realizable in artificial
platforms, such as
electrical circuits~\cite{lee2018}.  Since the limit of infinite uncorrelated
disorder gives rise to a trivial ultra-insulator, a competition between
different disorders should result in a transition between TUIs, as sketched in
Fig.~\ref{Fig:ultra_ins1}(c).  More extreme types of structural disorders, such
as amorphous crystals, have recently been shown to realize conventional
topological insulators~\cite{PhysRevLett.118.236402, Mitchell_2018,
Marsal_2020}, but the full topological classification of disordered electronic
matter is unknown, in particular extensions to disordered hyperuniform systems
\cite{PhysRevE.68.041113, doi:10.1126/sciadv.aba0826}.  Our work thus opens a
pathway to generalize notions of topology to wider range of systems.

\begin{acknowledgments}\textit{Acknowledgments} --- B.L. acknowledges financial
support from the Swiss National Science Foundation (Postdoc.Mobility Grant No.
214461). L.T.  acknowledges support by the FNS/SNF Ambizione Grant No.
PZ00P2\_179962.  T.N. acknowledges support from the Swiss National Science
Foundation through a Consolidator Grant (iTQC, TMCG-$2\_213805$) P.B.\ is
supported by the Deutsche Forschungsgemeinschaft (DFG) project Grant No.
277101999 within the CRC network TR 183 (subproject No. A03). 
\end{acknowledgments}

\bibliography{refs,ref}

\clearpage
\begin{widetext}
.
\setcounter{equation}{0}
\renewcommand\theequation{S\arabic{equation}}
\renewcommand\thesection{S\arabic{section}}

\section*{Supplementary Material}

\subsection{Classification of TUIs}

The key insight for the topological classification of ultra-localized insulators and ultra-localized superconductors (UIs) is that from the basis of localized eigenvectors $\ket{\psi_{\vn}}$ of an UI one may construct an auxiliary chiral Hamiltonian $\aH$, which has the unitary matrix $U_{\vr,\vn} = \langle \vr \ket{\psi_{\vn}}$ and its hermitian conjugate as its off-diagonal blocks, see Eq.\ (\ref{eq:tildeH}) of the main text. 
The auxiliary Hamiltonian is defined on a bipartite lattice $\mathcal{L}_{\vr} \cup \mathcal{L}_{\vn}$ consisting of the original lattice $\mathcal{L}$ (site labels $\vr$) and the lattice of eigenstate labels $\vn$, which is isomorphic to $\mathcal{L}$. With help of this insight, the topological classification of TUIs can be readily obtained from the classification of conventional topological phases in one of the three chiral classes AIII, BDI, or CII. 

A subtlety in this procedure is that the assignment of an auxiliary Hamiltonian $\aH$ to an ultra-insulator is not unique: $\aH$ is defined up to a local permutation of the eigenstate labels $\vn$ and up to a phase factor in each of the eigenstates $\ket{\psi_{\vn}}$. (The permutation of the eigenstate labels is local, because each eigenstate label $\vn$ must lie within the support of the localized eigenstate $\ket{\psi_{\vn}}$.) To account for the non-uniqueness of the auxiliary Hamiltonian we therefore not only need to determine the classifying group $K_{\rm aux}(d)$ of the auxiliary Hamiltonians in dimension $d$ --- i.e., their set of topological invariants with the corresponding group structure ---, but also the classifying group $K_{\rm aux}'(d)$ of auxiliary Hamiltonians that can be obtained from a trivial reference auxiliary Hamiltonian $\aH^0$ by local permutation of the site labels $\vn$ and/or by changing phases of eigenvectors $\ket{\psi_{\vn}}$. The classifying group $K^{\rm TUI}(d)$ of topological ultra-localized insulators and superconductors is then given by the quotient
\begin{equation}
    K^{\rm TUI}(d) = K_{\rm aux}(d)/K'_{\rm aux}(d).
\end{equation}

We will now implement this classification procedure for each of the ten Altland-Zirnbauer classes. Hereto, we will express $K_{\rm aux}(d)$ and $K_{\rm aux}'(d)$ in terms of the classifying groups ${\cal K}_{\rm AIII}(d)$, ${\cal K}_{\rm BDI}(d)$, and ${\cal K}_{\rm CII}(d)$ of conventional topological phases in the three chiral classes AIII, BDI, and CII.

\subsubsection{Wigner-Dyson classes A, AI, AII}
In this section, we provide details on the classification of TUIs in Wigner-Dyson classes. These are the three Altland-Zirnbauer classes that do not have an antisymmetry constraint (chiral or particle-hole).

{\em Class A.---} For TUIs in class A, no symmetry constraints are imposed on the eigenvector matrix $U$ (other than unitarity and locality), so that the auxiliary Hamiltonian \eqref{eq:tildeH} belongs to class AIII. Hence, for TUIs in class A we obtain
\begin{equation}
  K_{\rm aux}(d) = {\cal K}_{\rm AIII}(d),
\end{equation}
where ${\cal K}_{\rm AIII}(d)$ is the classifying group of conventional topological insulators in tenfold-way class AIII.

To find $K_{\rm aux}'(d)$, we note that the Hamiltonian $H$ of the UI determines the eigenstates $|\psi_{\vn} \rangle$ up to a phase factor $e^{i \phi_{\vn }}$ and an arbitrary relabeling of the indices $\vn$, as long as such a relabelling preserves the locality of $U$. The uncertainty of the phase factors means that the matrix $U$ may be right-multiplied by a diagonal matrix $D$ containing phase factors $e^{i \phi_{\vn }}$. The relabeling of the indices $\vn$ amounts to the degree of freedom to right-multiply $U$ by a ``relabeling matrix'' $P_{\vn',\vn}$. The relabeling matrix is a permutation matrix, {\em i.e.}, its entries are $0$ or $1$. Because of the condition that the site indices $\vn$ are chosen such that the eigenstate $|\psi_{\vn}\rangle$ is localized near lattice position $\vn$, the relabeling matrix is local in the site indices $\vn$ and $\vn'$, {\em i.e.}, $P_{\vn',\vn} = 0$ if $|\vn - \vn'|$ exceeds the localization length.
Auxiliary Hamiltonians that are obtained from the trivial reference Hamiltonian $\aH^0$ --- for which we take the eigenvector matrix $U^0 = \openone$ --- by permutation of the eigenstate labels or by changing the phases of the eigenstates $\ket{\psi_{\vn}}$ are of the form
\begin{equation}
  \aH'^0 = \begin{pmatrix} 0 & D P \\ P^{\dagger} D^{\dagger} & 0 \end{pmatrix}.
\end{equation}
Such a Hamiltonian describes a lattice model that consists of sites that are pairwise connected by a hopping bond of unit strength. Such dimer Hamiltonians can have a nonzero integer winding number for $d=1$ --- the Su–Schrieffer–Heeger model being the paradigmatic example ---, but not for $d > 1$.
Hence, we conclude that $K_{\rm aux}'(d) = {\cal K}_{\rm AIII}(1) = \ZZ$ if $d=1$ and $K_{\rm aux}'(d) = 0$ otherwise, so that
\begin{equation}
  K_{\rm A}^{\rm TUI}(d) = \left\{ \begin{array}{ll}
  0 & \mbox{if $d=1$}, \\
  {\cal K}_{\rm AIII}(d) & \mbox{if $d > 1$}. \end{array} \right.
\end{equation}

{\em Class AI.---} For class AI, $H$ is subject to the constraint
\begin{equation}
  H = H^*. \label{eq:HAI}
\end{equation}
Correspondingly, the matrix $U$ is an orthogonal matrix and the auxiliary Hamiltonian $\aH$ is in tenfold-way class BDI, so that $K_{\rm aux}(d) = {\cal K}_{\rm BDI}(d)$ for class AI.
In this case, the ``phase matrix'' $D$ consists of signs $\pm 1$. 
Since relabeling matrices $P$ may have arbitrary integer winding number in $d=1$, but possess no nontrivial topology if $d > 1$, it also follows that $K_{\rm aux}'(d) = {\cal K}_{\rm BDI}(1) = \ZZ$ if $d=1$ and $K_{\rm aux}'(d) = 0$ otherwise. We conclude that
\begin{equation}
  K_{\rm AI}^{\rm TUI}(d) = \left\{ \begin{array}{ll}
  0 & \mbox{if $d=1$}, \\
  {\cal K}_{\rm BDI}(d) & \mbox{if $d > 1$}. \end{array} \right.
\end{equation}

{\em Class AII.---} For class AII, $H$ must satisfy Eq.~\eqref{eq:timereversal}.
Eigenvalues of $H$ come in twofold (Kramers) degenerate pairs.
The same symmetry condition applies to the eigenvector matrix $U$, which makes it a symplectic matrix, and to the auxiliary Hamiltonian $\aH$, which places it  in tenfold-way class CII. Since the relabeling matrix $P$ is proportional to the $2 \times 2$ unit matrix $\sigma_0$, the corresponding winding numbers for $d=1$ are always even, so that $K_{\rm aux}'(d) = {\cal K}_{\rm CII}(1) = 2\ZZ$ if $d=1$ and $0$ otherwise. We then find
\begin{equation}
  K_{\rm AII}^{\rm TUI}(d) = \left\{ \begin{array}{ll}
  0 & \mbox{if $d=1$}, \\
  {\cal K}_{\rm CII}(d) & \mbox{if $d > 1$}. \end{array} \right.
\end{equation}
In $d=3$, class CII admits a $\ZZ_2$ classification, which carries over to the TUIs in class AII, as discussed in the main text.

\subsubsection{Chiral classes AIII, BDI, CII}

Hamiltonians in the chiral classes AIII, BDI, and CII are subject to an antisymmetry, which can be represented as
\begin{equation}
  H = - \tau_3 H \tau_3, \label{eq:Htau3}
\end{equation}
where $\tau_3$ is the third Pauli matrix in a suitably chosen local basis.
Further, in classes BDI and CII, $H$ is subject to an additional symmetry condition of the form $H = H^*$ or $H = \sigma_2 H^* \sigma_2$.
This additional symmetry condition commutes with $\tau_3$ and imposes that elements of $H$ are real or quaternion numbers.  
The antisymmetry condition (\ref{eq:Htau3}) also imposes that the eigenvalues come in pairs $\pm E_{\vn }$ and that the corresponding eigenvectors are related to each other by multiplication by $\tau_3$. In class CII, all eigenvalues are twofold (Kramers) degenerate.

In principle, an eigenvalue zero may appear. Such an eigenvalue is always twofold degenerate in classes AIII and BDI and fourfold degenerate in class CII. If this happens, it is important to distinguish an ``accidental'' zero crossing, for which the two localized eigenstates are related to each other by multiplication by $\tau_3$, and a ``topological'' zero-eigenvalue, for which either the eigenstates are not localized or, if they are localized, they are not related to each other by multiplication by $\tau_3$. For the bulk classification of TUIs we consider systems with periodic boundary conditions, in which anomalous zero-energy eigenstates  do not exist.

Without a topological eigenvalue at zero energy, one may then obtain a classification of TUIs in classes AIII, BDI, and CII by noting that, whereas the Hamiltonian $H$ anticommutes with $\tau_3$, the eigenvector matrix {\em commutes} with $\tau_3$, so that
\begin{equation}
  U = \mbox{diag}\, (U_+,U_-),
\end{equation}
where $U_+$ and $U_-$ are unitary, orthogonal, or symplectic matrices of equal size. 
The Hamiltonian $H$ then has the diagonal form \eqref{eq:TUIdefin}
where $\Lambda$ is a block-diagonal matrix containing blocks $E_{\vn } \tau_1$ for classes AIII and BDI and blocks $E_{\vn } \tau_1 \sigma_0$ for class CII, where each pair $\pm E_{\vn }$ of eigenvalues of $H$ appears once. (The eigenvalues $E_{\vn }$ appearing in $\Lambda$ may be positive or negative. Note that, strictly speaking, the matrix $U$ in this case cannot be called ``eigenvector matrix'', since $\Lambda$ is not strictly diagonal.) Right-multiplication of $U = \mbox{diag}\,(U_+,U_-)$ with a relabeling matrix $P$ proportional to $\tau_0$ for classes AIII and BDI and to $\tau_0 \sigma_0$ for class CII does not change $H$. Further, one may also right-multiply $U$ with a block-diagional ``phase matrix'' $D$. Its diagonal blocks $D_{\vn }$ consist of $\tau_0$ or $\tau_3$ times a phase factor, sign $\pm 1$, or a $2 \times 2$ matrix $\sigma \in \mbox{SU(2)}$ for classes AIII, BDI, and CII, respectively, while simultaneously changing the sign of the corresponding eigenvalue $E_{\vn }$ in the eigenvalue matrix $\Lambda$ if $D_{\vn }$ contains $\tau_3$.

Whereas there is no topological information in the diagonal matrix $\Lambda$, the two eigenvector matrices $U_{\pm}$ allow for the construction of a block-diagonal auxiliary Hamiltonian $\aH = \mbox{diag}(\aH^+,\aH^-)$. Each block contributes its own classifying group. Hence, for the chiral class X, where X stands for AIII, BDI, CII, we find
\begin{equation}
  K_{\rm aux}(d) = {\cal K}_{\rm X}(d) \times {\cal K}_{\rm X}(d).
\end{equation}
The ambiguities in the definitions of $U_{\pm}$ and $E$ can only produce topological phases corresponding to Hamiltonians with a dimerized or an on-site structure. Such Hamiltonians must have trivial topology for $d > 1$, so that $K_{\rm aux}'(d) = 0$ for $d > 1$. Since the relabeling matrix $P$ is proportional to $\tau_0$, the same relabeling applies to both $U_+$ and $U_-$. Hence, for $d=1$ the relabeling degree of freedom gives $K_{\rm aux}'(1) = {\cal K}_{\rm X}(1)$.
We thus find
\begin{equation}
  K_{\rm X}^{\rm TUI}(d) = \left\{ \begin{array}{ll}
  {\cal K}_{\rm X}(1) & \mbox{if $d=1$}, \\
  {\cal K}_{\rm X}(d)\times {\cal K}_{\rm X}(d) & \mbox{if $d > 1$},
  \end{array} \right.
\end{equation}
where $\rm X$ stands for AIII, BDI, or CII.

\subsubsection{Superconducting classes C, CI, D, DIII}

In the four superconducting classes C, CI, D, and DIII, $H$ satisfies an antiunitary antisymmetry and, for CI and DIII, an additional antiunitary symmetry. Classes  CI and DIII also have a chiral symmetry, the product of the antiunitary antisymmetry and symmetry, but, unlike for the chiral classes, time-reversal and particle-hole conjugation do not commute with the chiral conjugation. As we show below, this means that in all four superconducting classes, the eigenvectors of $H$ are characterized by a single unitary matrix only.

{\em Class D.---} In class D, the Hamiltonian $H$ satisfies an antiunitary antisymmetry
\begin{equation}
  H = - \tau_1 H^* \tau_1.
\end{equation}
Its eigenvalues come in pairs $\pm E_{\vn }$ and the eigenvectors are related to each other via particle-hole conjugation ({\em i.e.}, multiplication with $\tau_1$ and complex conjugation). For an ultra-localized Hamiltonian we impose the additional condition that this also applies to the eigenvalue $E = 0$, {\em i.e.}, we rule out the possibility of a ``topological zero-eigenvalue'' for which the eigenstates are localized, but not related by particle-hole conjugation.

The Hamiltonian $H$ has the eigenvalue decomposition
\begin{equation}
  H = U \Lambda U^{\dagger},
\end{equation}
where $\Lambda$ is a diagonal matrix containing blocks $E_{\vn} \tau_3$, with each pair $\pm E_{\vn}$ of eigenvalues of $H$ appearing once. The eigenvector matrix $U$ satisfies the antiunitary symmetry
\begin{equation}
  U = \tau_1 U^* \tau_1. \label{eq:UD}
\end{equation}
The eigenvector matrix can be right-multiplied by a block-diagonal matrix $D$ containing $2 \times 2$ blocks of the form $e^{i \phi_{\vn } \tau_3}$ or $\tau_1 e^{i \phi_{\vn } \tau_3}$ without changing $H$. Physically, this freedom corresponds to the freedom to choose the phases of the electron-like wavefunctions. The phases of the hole-like wavefunctions are then determined by particle-hole symmetry. If the matrix $\tau_1$ appears in a block, the corresponding element $E_{\vn }$ in the diagonal matrix $\Lambda$ must be multiplied by $-1$. 
Physically, this degree of freedom corresponds to an exchange of electron and hole labels, which must be accompanied by the change $E_{\vn } \to -E_{\vn }$. The relabeling matrix $P$ is proportional to $\tau_0$. 

The group of unitary matrices satisfying the property (\ref{eq:UD}) is isomorphic to the orthogonal group of matrices of the same size as $U$ and the basis transformation from unitary matrices satisfying Eq.\ (\ref{eq:UD}) to orthogonal matrices does not affect the locality condition. (Physically, this basis transformation corresponds to the change from the particle-hole basis to the Majorana basis.) Hence, the auxiliary Hamiltonian $\aH$ is in class BDI. For $d=1$, we note that $K_{\rm aux}'(1) = 2 \ZZ$ for class D, because winding numbers of the relabeling matrix $P$ are always even as $P$ is proportional to $\tau_0$.
Since $K_{\rm aux}(1) = {\cal K}_{\rm BDI}(1) = \ZZ$, we conclude that
\begin{equation}
  K_{\rm D}^{\rm TUI}(d) = \left\{ \begin{array}{ll}
  \ZZ_2 & \mbox{if $d=1$}, \\
  {\cal K}_{\rm BDI}(d) & \mbox{if $d > 1$}.
  \end{array} \right.
\end{equation}
The nontrivial topological phase for $d=1$ has zero-energy Majorana modes at its ends. (There are no Majorana zero modes if periodic boundary conditions are applied.) The zero-energy Majorana modes that appear for open boundary conditions count as topological zero modes, because they are not related to each other by particle-hole conjugation.

{\em Class C.---} In tenfold-way class C, the Hamiltonian $H$ satisfies the antiunitary antisymmetry
\begin{equation}
  H = - \tau_2 H^* \tau_2.
\end{equation}
As in class D, eigenvalues come in pairs $\pm E_{\vn }$ and one may write
\begin{equation}
  H = U \Lambda U^{\dagger},
  \label{eq:UC}
\end{equation}
where $U = \tau_2 U^* \tau_2$
is a symplectic matrix and $\Lambda$ is a block-diagional matrix containing blocks $E_{\vn } \tau_3$, where each eigenvalue pair $\pm E_{\vn }$ appears once. The Hamiltonian $H$ is not changed if $U$ is right-multiplied by a block-diagonal matrix $D$ containing blocks of the form $\tau_0 e^{i \phi_{\vn } \tau_3}$ or $\tau_2 e^{i \phi_{\vn } \tau_3}$, where, in the latter case, $E_{\vn }$ must be simultaneously multiplied by $-1$. The unitary matrix $U$ may also be right-multiplied by a local relabeling matrix $P$, which is proportional to $\tau_0$. Class C does not admit topological zero modes.

Since the eigenvector matrix $U$ is symplectic for class C, see Eq.\ (\ref{eq:UC}), the auxiliary Hamiltonian $\aH$ is in tenfold-way class CII. As winding numbers in class CII are always even for $d=1$, one finds that $K_{\rm aux}'(1) = {\cal K}_{\rm CII}(1) = 2 \ZZ$. Hence
\begin{equation}
  K_{\rm C}^{\rm TUI}(d) = \left\{ \begin{array}{ll}
  0 & \mbox{if $d = 1$}, \\
  {\cal K}_{\rm CII}(d) & \mbox{if $d > 1$}.
  \end{array} \right.
\end{equation}
We note that a topological ultra-localized superconductor with a $\ZZ_2$ classification exists for $d=3$. Since ${\cal K}_{\rm C}(3) = 0$, this phase must be trivial in the conventional sense.

{\em Class CI.---} Hamiltonians in class CI satisfy the chiral symmetry
\begin{equation}
  H = - \tau_3 H \tau_3
\end{equation}
and the particle-hole antisymmetry
\begin{equation}
  H = -\tau_2 H^* \tau_2. 
\end{equation}
The eigenvalues come in pairs $\pm E_{\vn}$ and one may write
\begin{equation}
  H = U \Lambda U^{\dagger}, \quad U = \mbox{diag}\, (U_+,U_-), \quad U_- = U_+^*,
\end{equation}
where $\Lambda$ is a block-diagional matrix containing blocks $E_{\vn } \tau_1$, such that each eigenvalue pair $\pm E_{\vn }$ appears once. The matrix $U_+$ is a unitary matrix. The Hamiltonian $H$ remains unchanged if $U$ is right-multiplied by a block-diagonal ``phase matrix'' $D$ having $2 \times 2$ diagonal blocks $D_{\vn }$ equal to $\pm \tau_0$ or $\pm i \tau_3$, whereby $E_{\vn }$ is simultaneously multiplied by $-1$ if $D_{\vn } = i \tau_3$ or $-i \tau_3$. The unitary matrix $U$ may also be right-multiplied by a local relabeling matrix $P$, which is proportional to $\tau_0$. Class CI has no topological zero modes.

To construct the auxiliary Hamiltonian, one uses the matrix $U_+$ only. Since $U_+$ is a unitary matrix without further symmetries, the auxiliary Hamiltonian $\aH$ is in tenfold-way class AIII. We note that $H$ is unchanged if $U_+$ is right-multiplied by a diagonal matrix $D_+$ with diagonal elements $D_{+,\vn }$ equal $\pm 1$ or $\pm i$, whereby one has to simultaneously multiply $E_{\vn }$ by $-1$ if $D_{+,\vn } = \pm i$. Since a relabeling matrix $P$ is proportional to $\tau_0$, one may define a relabeling matrix $P_+$ that acts on $U_+$ only. In $d=1$ one has $K_{\rm aux}'(1) = {\cal K}_{\rm AIII}(1) = \ZZ$, whereas $K_{\rm aux}'(1) = 0$ for $d > 1$. It follows that
\begin{equation}
  K_{\rm CI}^{\rm TUI}(d) = \left\{ \begin{array}{ll}
  0 & \mbox{if $d = 1$}, \\
  {\cal K}_{\rm AIII}(d) & \mbox{if $d > 1$}.
  \end{array} \right.
\end{equation}
We thus find that TUIs in class CI exists for all odd dimensions $d>1$ and that they have an integer classification. 

{\em Class DIII.---} In class DIII, $H$ satisfies the chiral symmetry $H = - \tau_3 H \tau_3$ 
and the particle-hole antisymmetry
 $H = -\tau_2 \sigma_2 H^* \tau_2 \sigma_2$. 
The eigenvalues come in twofold degenerate (Kramers) pairs $\pm E_{\vn}$ and one may write
\begin{equation}
  H = U \Lambda U^{\dagger}, \quad U = \mbox{diag}\, (U_+,U_-), \quad U_- = \sigma_2 U_+^* \sigma_2.
\end{equation}
Here $\Lambda$ is a block-diagonal matrix containing blocks $E_{\vn } \tau_1 \sigma_0$, such that each Kramers pair $\pm E_{\vn }$ appears once. The matrix $U_+$ is a unitary matrix without further symmetries imposed. The Hamiltonian $H$ remains unchanged if $U$ is right-multiplied by a block-diagonal ``phase matrix'' $D$ having $4 \times 4$ diagonal blocks $D_{\vn }$ equal to $\pm \tau_0 \sigma_{\vn }$ or $\pm i \tau_3 \sigma_{\vn }$ with $\sigma_{\vn } \in \mbox{SU(2)}$, whereby $E_{\vn }$ is simultaneously multiplied by $-1$ if $D_{\vn }$ contains $\tau_3$. The unitary matrix $U$ may also be right-multiplied by a local relabeling matrix $P$, which is proportional to $\tau_0 \sigma_0$. Class DIII admits twofold (Kramers) degenerate topological zero modes. Hamiltonians $H$ with such topological zero-energy Kramers pairs must be excluded when defining topological ultra-localized superconductor in class DIII.

As in class CI, to construct the auxiliary Hamiltonian, one uses the matrix $U_+$ only. Since no symmetries are imposed on $U_+$, the auxiliary Hamiltonian $\aH$ is in tenfold-way class AIII. We note that $H$ is unchanged if $U_+$ is right-multiplied by a diagonal matrix $D_+$ with $2 \times 2$ diagonal blocks $D_{+,\vn }$ equal $\pm \sigma_{\vn }$ or $\pm i \sigma_{\vn }$, whereby one has to simultaneously multiply $E_{\vn }$ by $-1$ if $D_{+,\vn }$ contains a factor $i$. Since a relabeling matrix $P$ is proportional to $\tau_0 \sigma_0$, one may define a relabeling matrix $P_+$ proportional to $\sigma_0$ that acts on $U_+$ only. Since $P_+$ is proprortional to $\sigma_0$, in $d=1$ one has $K_{\rm aux}'(1) = 2 \ZZ$, whereas ${\cal K}_{\rm AIII} = \ZZ$, so that
\begin{equation}
  K_{\rm DIII}^{\rm TUI}(d) = \left\{ \begin{array}{ll}
  \ZZ_2 & \mbox{if $d=1$}, \\
  K_{\rm AIII}(d) & \mbox{if $d > 1$}.
  \end{array} \right.
\end{equation}
We thus find that TUIs in class DIII exists for all odd dimensions $d>1$ and that they have an integer classification. For $d=1$, localized superconductors in class DIII have a $\ZZ_2$ classification, which coincides with the conventional classification of topological superconductors in that class in that dimension. The nontrivial topological ultra-localized superconductor for $d=1$ has Majorana Kramers pairs as its anomalous end states.

\subsubsection{Classification table}

The classification procedure outlined above results in the classification of Table \ref{tab:classification_TUIs} from the main text for $1 \le d \le 9$. For dimensions $d > 9$ the classification repeats itself with a period-eight Bott periodicity starting from $d=2$. The classification of TUIs for $d=1$ deviates from the period-eight pattern. It is the same as the classification of conventional topological phases. Indeed, for $d=1$ infinitesimal disorder turns each topological insulator or superconductor into a TUI, so that there is a one-to-one correspondence between conventional topological phases for $d=1$ and topological phases of ultra-localized insulators and superconductors.

\subsection{Topological invariants}

\subsubsection{Conventional topological insulators and superconductors}

\begin{table}[!]
\begin{center}
	\begin{tabular}{l@{\extracolsep{\fill}} ccccccccc} \hline\hline
                                class  & $\,d=1$ & $d=2$ & $d=3$ & $d=4$ & $d=5$ & $d=6$ & $d=7$ & $d=8$ & $d=9$\\ \hline
				A      & $0$ & $\Ch{1}$ & $0$ & $\Ch{2}$ & $0$ & $\Ch{3}$ & $0$ & $\Ch{4}$ & $0$\\
				AIII   & $\W{1}$ & $0$  & $\W{3}$ & $0$ & $\W{5}$ & $0$ & $\W{7}$ & $0$ & $\W{9}$\\
                                \hline
				AI     & $0$ & $0$ & $0$ &$\overbar{\Ch{2}}$ & $0$ & $\beta_\Ch{3}$ & $\alpha_\Ch{3}$ & $\Ch{3}$ & $0$\\
				BDI    & $\W{1}$  & $0$ & $0$ & $0$ & $\overbar{\W{5}}$ & $0$ & $\beta_\W{9}$ & $\alpha_\W{9}$ & $\W{9}$\\
				D      & $\alpha_\Ch{1}$ &  $\Ch{1}$ & $0$ & $0$ & $0$ & $\overbar{\Ch{3}}$ & $0$ & $\beta_\Ch{5}$ & $\alpha_\Ch{5}$\\
				DIII   & $\beta_\W{3}$ & $\alpha_\W{3}$ & $\W{3}$ & $0$ & $0$ & $0$ & $\overbar{\W{7}}$ & $0$ & $\beta_\W{11}$\\
				AII    & $0$ & $\beta_\Ch{2}$ & $\alpha_\Ch{2}$ & $\Ch{2}$ & $0$ & $0$ & $0$ & $\overbar{\Ch{4}}$ & $0$\\
				CII    & $\overbar{\W{1}}$ & $0$ & $\beta_\W{5}$ & $\alpha_\W{5}$ & $\W{5}$ & $0$ & $0$ & $0$ & $\overbar{\W{9}}$\\
				C   & $0$ & $\overbar{\Ch{1}}$ & $0$ & $\beta_\Ch{3}$ & $\alpha_\Ch{3}$ & $\Ch{3}$ & $0$ & $0$ & $0$\\
				CI   & $0$ & $0$ & $\overbar{\W{3}}$ & $0$ & $\beta_\W{7}$ & $\alpha_\W{7}$ & $\W{7}$ & $0$ & $0$\\
                                \hline\hline
                        \end{tabular}
			\caption{Topological invariants of tenfold-way
				topological phases. The integer invariants are Chern or winding numbers or higher-dimensional generalizations thereof. The $\ZZ_2$ invariants are
				denoted by $\alpha_X$ ($\beta_X$) for first
				(second) descendant of the integer invariant X. A ``2'' in front of the integer invariant indicates that the invariant takes  even
				values only.\label{tab:TFinv}}
	\end{center}
\end{table}

The topological invariants of conventional topological insulators and superconductors are well known \cite{ryu2010}. They are summarized in Table~\ref{tab:TFinv}.
The topological invariants are of integer ($\ZZ$) or of even-odd ($\ZZ_2$) type. The integer invariant is the Chern number $\Ch{n}$ or a higher-dimensional generalization of it for even dimensions $d=2n$ or it is the winding number $\W{2n-1}$ or a higher-dimensional generalization for odd dimensions $d=2n-1$. 
For the complex classes A and AIII the topological invariant is always an integer. For the real classes, there is one series of integer invariants which can be even or odd and one series of integer invariants that can take even values only. 
In addition, there are two series of $\ZZ_2$-type descendant invariants, which derive from the former series of integer invariants.
The first (second) descendant $\alpha_X$ ($\beta_X$) is 
equal to the parity of the parent integer invariant $X$ in the same
tenfold-way class, but in one (two) spatial dimension(s) higher, see
Tab.~\ref{tab:TFinv}.

\subsubsection{Topological ultra-insulators and superconductors}

In dimensions $d > 1$ the topological invariants for TUIs are the topological invariants associated with the auxiliary Hamiltonian $\aH$. Since $\aH$ is chiral, the TUI invariant is of the same type as the invariant for a conventional topological phase in one of the chiral classes AIII, BDI, or CII, or a pair of such invariants. Table~\ref{tab:TUIinv} gives an overview of these invariants.
For $d=1$ the classification of TUIs is the same as that of conventional topological phases, so that the topologial invariants of the conventional classification can be used. These invariants can be taken directly from Tab.~\ref{tab:TFinv}.

\begin{table}[h!]
\begin{center}
\begin{tabular}[t]{l   @{\extracolsep{\fill}} |c|cccccccc}
\hline\hline 
				class   & $d=1$ & $d=2$                              & $d=3$                                  & $d=4$                                      & $d=5$ & $d=6$ & $d=7$ & $d=8$ & $d=9$\\ \hline
				AIII      & $\W{1}$ & $0$       & $(\W{3},\W{3}^\prime)$           & $0$               & $(\W{5},\W{5}^\prime)$ & $0$ & $(\W{7},\W{7}^\prime)$ & $0$ & $(\W{9},\W{9}^\prime)$      \\
				A       & $0$ & $0$       & $\W{3}$           & $0$               & $\W{5}$ & $0$ & $\W{7}$ & $0$ & $\W{9}$      \\
				DIII  & $\beta_\W{3}$ & $0$  & $\W{3}$ & $0$ & $\W{5}$ & $0$ & $\W{7}$ & $0$ & $\W{9}$\\
				CI   & $0$ & $0$  & $\W{3}$ & $0$ & $\W{5}$ & $0$ & $\W{7}$ & $0$ & $\W{9}$\\
				\hline
				BDI & $\W{1}$  & $0$ & $0$ & $0$ & $(\overbar{\W{5}},\overbar{\W{5}}^\prime)$ & $0$ & $(\beta_\W{9},\beta_\W{9}^\prime)$ & $(\alpha_\W{9},\alpha_\W{9}^\prime)$ & $(\W{9},\W{9}^\prime)$\\
				AI  & $0$  & $0$ & $0$ & $0$ & $\overbar{\W{5}}$ & $0$ & $\beta_\W{9}$ & $\alpha_\W{9}$ & $\W{9}$\\
				D   & $\alpha_\Ch{1}$  & $0$ & $0$ & $0$ & $\overbar{\W{5}}$ & $0$ & $\beta_\W{9}$ & $\alpha_\W{9}$ & $\W{9}$\\
				\hline
				CII  & $\overbar{\W{1}}$ & $0$ & $(\beta_\W{5},\beta_\W{5}^\prime)$ & $(\alpha_\W{5},\alpha_\W{5}^\prime)$ & $(\W{5},\W{5}^\prime)$ & $0$ & $0$ & $0$ & $(\overbar{\W{9}},\overbar{\W{9}}^\prime)$\\
				AII  & $0$ & $0$ & $\beta_\W{5}$ & $\alpha_\W{5}$ & $\W{5}$ & $0$ & $0$ & $0$ & $\overbar{\W{9}}$\\
				C  & $0$ & $0$ & $\beta_\W{5}$ & $\alpha_\W{5}$ & $\W{5}$ & $0$ & $0$ & $0$ & $\overbar{\W{9}}$\\
				\hline\hline
			\end{tabular}
			\caption{Topological invariants clasifying TUI phases. The ten Altland-Zirnbauer classes are
				organized into three groups, corresponding to the chiral class of the associated Hamiltonian $\aH$. Accordingly, the topological invariants are derived from those in classes AIII, BDI, and CII, respectively, see
				Table~\ref{tab:TFinv}. The classification for $d=1$ is an exception to this rule and follows the classification of conventional topological phases in one dimension.}

			\label{tab:TUIinv}
	\end{center}
\end{table}

\subsection{Relation between TUI classification and TI classification}

\subsubsection{Topological invariants}

Every ultra-localized insulator or superconductor is also an insulator or superconductor according to the conventional definition. In the same way, topological equivalence as ultra-localized phases also implies topological equivalence as a conventional topological phase. This implies that there is a group homomorphism
\begin{equation}
  f:\, K^{\rm TUI}(d) \to {\cal K}(d) \label{eq:f}
\end{equation}
that assigns to each topological class of the TUI classification the conventional tenfold-way topological class.  
Table \ref{tab:TF-TUI} summarizes this relation for all Altland-Zirnbauer symmetry classes and for all dimensions. A derivation will be given at the end of this Section.

\begin{table*}[!]
\begin{center}
	\begin{tabular}{l@{\extracolsep{\fill}} |c|cccccccc} \hline\hline
                                class  & $\,d=1$ & $d=2$ & $d=3$ & $d=4$ & $d=5$ & $d=6$ & $d=7$ & $d=8$ & $d=9$\\ \hline
				A      & $0$ & $\xmark$ & $0$ & $\xmark$ & $0$ & $\xmark$ & $0$ & $\xmark$ & $0$\\
				AIII   & $\W{1}$ & $0$  & $\W{3}^\prime-\W{3}$ & $0$ & $\W{5}^\prime-\W{5}$ & $0$ & $\W{7}^\prime-\W{7}$ & $0$ & $\W{9}^\prime-\W{9}$\\
                                \hline
				AI     & $0$ & $0$ & $0$ &$\xmark$ & $0$ & $\xmark$ & $\xmark$ & $\xmark$ & $0$\\
				BDI    & $\W{1}$  & $0$ & $0$ & $0$ & $\overbar{\W{5}}^\prime-\overbar{\W{5}}$ & $0$ & $\beta_\W{9}^\prime+\beta_\W{9}$ & $\alpha_\W{9}^\prime-\alpha_\W{9}$ & $\W{9}^\prime-\W{9}$\\
				D      & $\alpha_\Ch{1}$ &  $\xmark$ & $0$ & $0$ & $0$ & $\xmark$ & $0$ & $\alpha_\W{9}$ & $\W{9}\mod 2$\\
				DIII   & $\beta_\W{3}$ & $\xmark$ & $2\W{3}$ & $0$ & $0$ & $0$ & $2\W{7}$ & $0$ & $\W{9}\mod 2$\\
				AII    & $0$ & $\xmark$ & $\xmark$ & $\xmark$ & $0$ & $0$ & $0$ & $\xmark$ & $0$\\
				CII    & $\overbar{\W{1}}$ & $0$ & $\beta_\W{5}^\prime-\beta_\W{5}$ & $\alpha_\W{5}^\prime-\alpha_\W{5}$ & $\W{5}^\prime-\W{5}$ & $0$ & $0$ & $0$ & $\overbar{\W{9}}^\prime-\overbar{\W{9}}$\\
				C   & $0$ & $\xmark$ & $0$ & $\alpha_\W{5}$ & $\W{5}\mod 2$ & $\xmark$ & $0$ & $0$ & $0$\\
				CI   & $0$ & $0$ & $2\W{3}$ & $0$ & $\W{5}\mod 2$ & $\xmark$ & $2\W{7}$ & $0$ & $0$\\
                                \hline\hline
                        \end{tabular}
			\caption{Conventional topological invariants of ultra-localized topological phases, expressed in terms of the invariants of the ultra-localized phases of Table \ref{tab:TUIinv}. Each entry in the table shows the image of the TUI invariant of Table \ref{tab:TUIinv} under the group homomorphism $f$ of Eq.\ (\ref{eq:f}). }
	\label{tab:TF-TUI}
	\end{center}
\end{table*}

\subsubsection{Wannier localizability of conventional tenfold-way classes}

The image of the map $f$ consists of those conventional topological insulators and superconductors that can be continuously deformed to an ultra-localized limit, {\em i.e.}, that admit a basis of exponentially localized eigenstates. Such topological phases are called {\em Wannier localizable} \cite{PhysRevX.14.011057}. Conversely, conventional topological phases that are not in the image of $f$ are non-localizable. In those insulators and superconductors there exists a topological obstruction to exponential localization of all eigenstates. Table \ref{tab:TF} of the main text lists which topological classes of the conventional tenfold-way classification are Wannier localizable and which ones are not. Examples of non-localizable phases are the quantum Hall and quantum spin-Hall insulators and the three-dimensional topological insulator. An example of a Wannier localizable topological phase is the one-dimensional topological superconductor \cite{PhysRevLett.113.046802}.

In a topological phase diagram, the phase of a non-localizable TI is enclosed by a line or a band of critical states, as shown schematically in Fig.~\ref{Fig:loc_nonloc}a, and topological properties disappear or change when approaching the transition to an ultra-localized insulator. 
On other hand, for Wannier localizable TIs there exists a parameter that drives the cross-over to an ultra-insulator without affecting the topology of the phase, see Fig.~\ref{Fig:loc_nonloc}b. In other words: The topology of non-localizable TIs is incompatible with being an ultra-insulator, whereas Wannier localizable TIs can be continuously deformed to an ultra-localized phase.

\begin{figure}[t]
\includegraphics[width=0.4\textwidth]{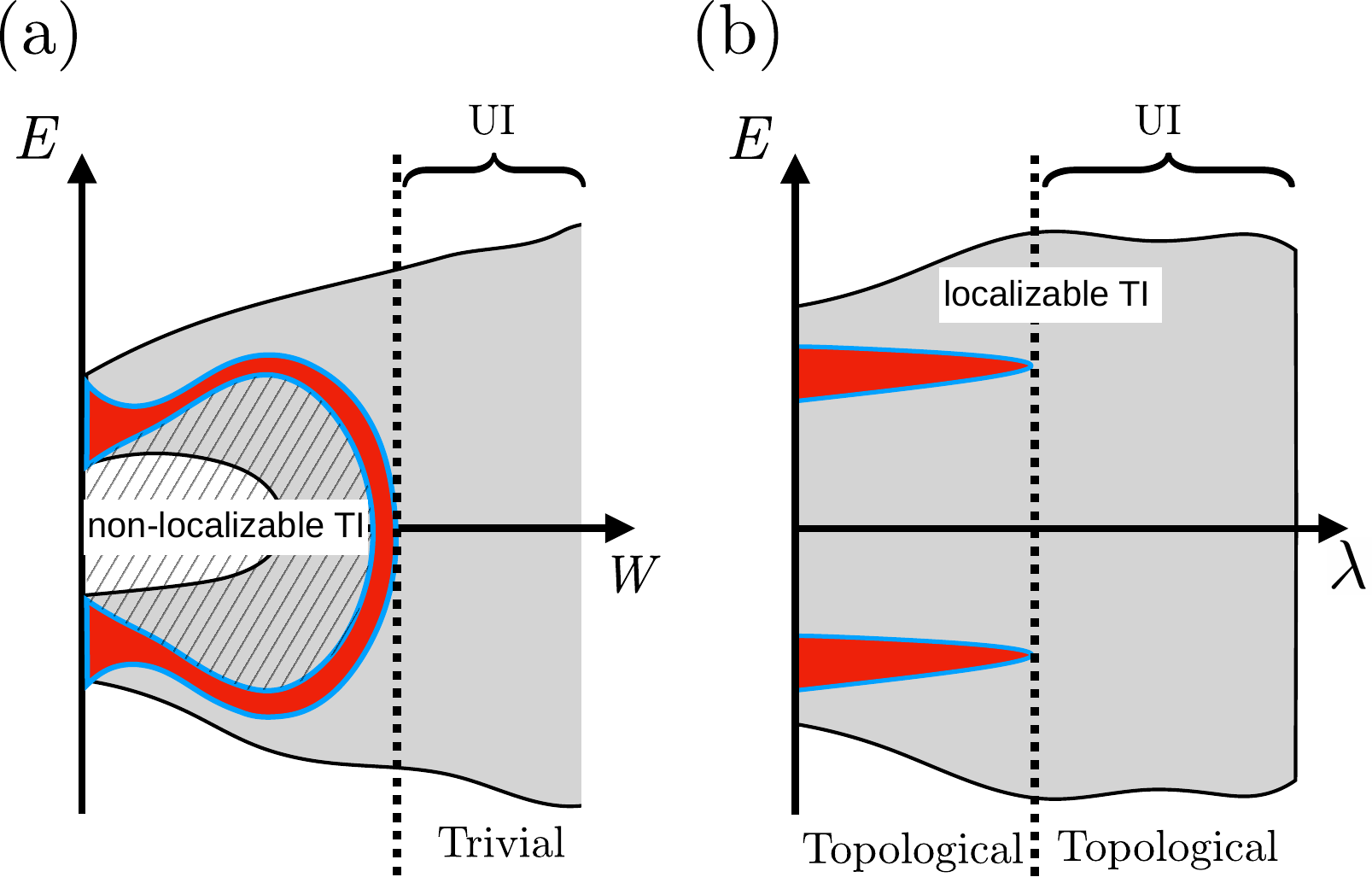}
 \caption{Conventional topological phases come in two flavors: (a) Non-localizable TIs, such as the quantum Hall or quantum spin Hall insulator, necessarily have extended bulk states below and above the Fermi level. Disappearance of these extended states is always accompanied by a transition to a different (usually trivial) phase. (b) Wannier-localizable TIs may have extended states above or below the Fermi level, but these states can be localized by varying some parameter $\lambda$ without a phase transition, i.e., the band of extended states does not enclose the topological phase.
 }
\label{Fig:loc_nonloc}
\end{figure}

\subsubsection{Mapping between topological invariants}

We now derive the relation between topological invariants according to the TUI classification and the topological invariants according to the conventional tenfold-way clasification that is shown in Table \ref{tab:TF-TUI}.

{\em Wigner-Dyson classes A, AI, and AII.---} 
Conventional topological phases in the Wigner-Dyson classes A, AI, and AII are always non-localizable \cite{PhysRevX.14.011057}. Hence the image of the group homomorphism $f$ is zero for the symmetry classes A, AI, and AII for all dimensions.

{\em Chiral clases AIII, BDI, and CII.---}
To find the relation between the invariants according to the TUI classification and the conventional classification, we recall that the eigenvector matrix $U$ for the chiral classes is of the form $U = \mbox{diag}\, (U_+,U_-)$. To determine the topological invariants, we may set all eigenvalues $E_n$ to $1$, so that the Hamiltonian $H$ becomes
\begin{equation}
  H = \begin{pmatrix} 0 & U_+ U_-^{\dagger} \\ U_- U_+^{\dagger} & 0 \end{pmatrix}.
\end{equation}
Its tenfold-way topological invariant is the difference of the tenfold-way invariants associated with $U_+$ and $U_-$.

{\em Superconducting classes C, CI, D, DIII.---}
In order to draw conclusions for the superconducting classes, we relate them to the Wigner-Dyson or chiral classes by either ``forgetting'' a symmetry or by imposing it by  ``doubling'', as we now explain.

\textit{Forgetting a symmetry} --- We start by proving the following statement: topologically
non-trivial superconductors in class C (D) in dimensions $d=4n+1$ and $d=4n$,
with $n$ an odd (even) integer, can be obtained by forgetting the time-reversal
symmetry of a generating topological superconductor in class CII (BDI).
Furthermore, topologically non-trivial insulators in class AII (AI) in
dimensions $d=4n-1$, with $n$ odd (even) integer, can be obtained by forgetting
the particle-hole symmetry of a generating topological superconductor in class
DIII (CI). To prove these statements, it is sufficient to consider models for generators of the topological phases with the additional time-reversal symmetry or particle-hole symmetry and show that these remain generators of the topological phases if this additional symmetry constraint is lifted.

We start from a $(4n+1)$-dimensional topologically non-trivial superconductor
in class CII (BDI) for $n$ odd (even), with the winding number $\W{}=1$. The
Hamiltonian of such system can be expressed using $4n+2$ (anticommuting) Dirac matrices,
\begin{align}
	(\Gamma_0,\dots,\Gamma_{4n+1}),
	\label{eq:1}
\end{align}
where chiral and time-reversal symmetry are represented by ${\cal C}=U_{\cal
C}$ and ${\cal T}=U_{\cal T}K$, respectively, with unitary matrices $U_{\cal C}$ and $U_{\cal T}$. The first descendent of the above
Hamiltonian,
\begin{align}
	(\Gamma_0,\dots,\Gamma_{4n}),
	\label{eq:2}
\end{align}
is non-trivial because $\W{}$ is odd. Furthermore, the phase~(\ref{eq:1}) is
non-trivial, when seen as a class-AIII phase. Via the dimension-increasing map, see Ref.\ \cite{teo2014}, one may obtain the Dirac matrices for a topological phase one dimension higher and in a symmetry class shifted by one according to the Bott clock: Starting from the Dirac matrices (\ref{eq:1}), we find that the Dirac matrices
\begin{align}
	(\Gamma_0,\dots,\Gamma_{4n+1},U_{\cal C}),
	\label{eq:3}
\end{align}
represent a generator of a $(4n+2)$-dimensional topological phase with Chern number $\text{Ch}=1$. Since $U_{\cal C}$
breaks time-reversal symmetry, but not particle-hole symmetry, the above system belongs to class C (D) for
$n$ odd (even). In this case, the two descendants exist, i.e., the sets of Dirac matrices,
\begin{align}
	(\Gamma_0,\dots,\Gamma_{4n+1}),
	\label{eq:4}
\end{align}
and
\begin{align}
	(\Gamma_0,\dots,\Gamma_{4n}),
	\label{eq:5}
\end{align}
both represent generators of  topologically non-trivial since $\text{Ch}$ is odd. Since
the phases characterized by the Dirac matrices~(\ref{eq:4}) and~(\ref{eq:5}) do not have the time-reversal symmetry
constraint, but are otherwise the same as (\ref{eq:1}) and~(\ref{eq:2}),
respectively, we have proven the first statement. 

Next, we consider the generator of a $(4n-1)$-dimensional topological superconducting phase in class DIII
(CI) for $n$ odd (even), which has winding number $\W{}=1$. This generator can be expressed
using the Dirac matrices
\begin{align}
	(\Gamma_0,\dots,\Gamma_{4n-1}).
	\label{eq:5a}
\end{align}
After applying the dimension-increasing map, we obtain the Dirac matrices
\begin{align}
	(\Gamma_0,\dots,\Gamma_{4n-1},U_{\cal C}),
	\label{eq:6a}
\end{align}
which describe a $4n$-dimensional topological insulator in class AII (AI) with (generalized) Chern number
$\text{Ch}=1$ for $n$ odd (even). Its first descendant is the $(4n-1)$-dimensional
topological insulator in the same class
\begin{align}
	(\Gamma_0,\dots,\Gamma_{4n-1}),
	\label{eq:7a}
\end{align}
which is the same as Eq.~(\ref{eq:5a}). This proves the second statement.

\textit{Doubling } ---
We now prove the following statement: All topologically non-trivial
superconductors in class CI (DIII) in dimensions $d=4n+2$ and $d=4n+1$, with $n$ an
odd (even) integer, can be obtained from the direct sum of a topological superconductor in class C (D) and its time reversed. We refer to this procedure as ``doubling''. Again, it is sufficient to prove this statement for generators of the nontrivial topology.

We start from a $(4n+2)$-dimensional topologically non-trivial superconductor
in class C (D) for $n$ odd (even), with Chern number $\text{Ch}=1$. The
Hamiltonian of such system can be expressed using $4n+3$ Dirac matrices,
\begin{align}
	(\Gamma_0,\dots,\Gamma_{4n+2}),
	\label{eq:6}
\end{align}
where particle-hole symmetry is represented by ${\cal P}=U_{\cal P}K$. The first
descendent of the above Hamiltonian,
\begin{align}
	(\Gamma_0,\dots,\Gamma_{4n+1}),
	\label{eq:7}
\end{align}
is non-trivial because $\text{Ch}$ is odd.

The phase~(\ref{eq:6}) is topologically non-trivial when seen as a class-A phase. Hence, by the dimension-increasing map, we conclude that,
\begin{align}
	(\tau_1\Gamma_0,\dots,\tau_1\Gamma_{4n+2},\tau_2),
	\label{eq:8}
\end{align}
represents a $(4n+3)$-dimensional system with an additional chiral symmetry $U_{\cal
C}=\tau_3$ and winding number  $\W{}=1$. Since the above phase respects particle-hole symmetry
${\cal P}=U_{\cal P}K$, it belongs to class CI (DIII) for $n$ odd (even). In
this case, the two descendants exist, i.e., the systems described by the Dirac matrices
\begin{align}
	(\tau_1\Gamma_0,\dots,\tau_1\Gamma_{4n+2})
	\label{eq:9}
\end{align}
and
\begin{align}
	(\tau_1\Gamma_0,\dots,\tau_1\Gamma_{4n+1})
	\label{eq:10}
\end{align}
are topologically non-trivial in the corresponding Altland-Zirnbauer classes. The
systems~(\ref{eq:8}) and~(\ref{eq:10}) are obtained by doubling the
systems~(\ref{eq:6}) and~(\ref{eq:7}), respectively, in order to introduce
chiral (or equivalently time-reversal) symmetry.

\textit{Implications for Wannier localizability versus non-localizability and for the group homomorphism $f$.} ---
Topological superconductors with $d=4,5$ ($d=8,9$) in class C
(D) can be obtained by forgetting the time-reversal symmetry of a topologically non-trivial phase in chiral class CII (BDI). Since topological phases in the chiral classes are always localizable, these phases are also Wannier localizable. Topological
superconductors with $d=5$ ($d=9$) in class CI (DIII) can be obtained by
doubling a phase in class C (D) in the same dimension, which we just found to localizable. Therefore, these two phases are localizable, too. Wannier localizabilty implies that the group homomorphism $f$ between the TUI and conventional classifications is surjective. Since the classifying groups are the same for these classes, $f$ must be the identity, as indicated in Table \ref{tab:TF-TUI}.

Finally, the remaining topological superconductors in $d=3,7$ all have the winding number as their
topological invariant. It is straightforward to see that they can be obtained by doubling a
corresponding phase from class AIII. This way, we obtain the localizability of
topological superconductors with even winding numbers in classes DIII and CI. 
For odd topological invariants in class DIII (CI), we note that forgetting
particle-hole symmetry gives nontrivial topological insulators in class AII (AI), which are
not localizable. Hence, $(4n-1)$-dimensional topological superconductors in
class DIII (CI), with $n$ odd (even), are non-localizable if their winding
number is an odd integer. For these classes, the group homomorphism $f$ is uniquely determined by requiring that its image consists of even winding numbers only, see Table \ref{tab:TF-TUI}.

\subsection{Lattice model for three-dimensional TUI in class AII}

Our starting point for the construction of a lattice model of a TUI in class AII in three dimensions is the eight-band lattice model $H_{\rm CII}$ of Eq.\ (\ref{eq:HCII}). This Hamiltonian may be diagonalized as
\begin{equation}
  H_{\rm CII}(\vk) = V(\vk) (\lambda(\vk) \tau_1) V(\vk)^{\dagger},
\end{equation}
where $V(\vk) = \mbox{diag}\,(V_+(\vk),V_-(\vk))$ is an $8 \times 8$ unitary matrix and $\lambda(\vk)$ is a $4\times 4$ diagonal matrix with positive elements $\lambda_j(\vk)$, $j=1,2,3,4$. The auxiliary Hamiltonian is obtained by flattening the spectrum of $H_{\rm CII}(\vk)$, {\em i.e.}, by setting $\lambda_j(\vk) = 1$, $j=1,2,3,4$. This gives
\begin{align}
  \aH(\vk) =&\, V(\vk) \tau_1 V^{\dagger}(\vk) \nonumber \\ =&\,
  \begin{pmatrix} 0 & V_+(\vk) V_-(\vk)^{\dagger} \\ V_-(\vk) V_+(\vk)^{\dagger} & 0 \end{pmatrix}.
\end{align} 
The Fourier transform $U_{\vr,\vn} = U_{\vr - \vn}$ of the unitary matrix $U(\vk) = V_+(\vk) V_-(\vk)^{\dagger}$ is the eigenvector matrix of the sought TUI in class AII. Since the auxiliary Hamiltonian $\aH(\vk)$ is a smooth function of $\vk$, $U_{\vr,\vn}$ is local. The eigenvector matrix $U_{\vr,\vn}$ defines the eigenstates $\ket{\psi_{\vn}} = \sum_{\vr} U_{\vr,\vn} \ket{\vr}$ of the TUI Hamiltonian $H$.

We consider the parameter values $v = 2$, $w=4$, $\lambda = 0.3$ in Eq.~\eqref{eq:HCII}. For these parameter values, $H_{\rm CII}(\vk)$ is topologically nontrivial in the conventional sense. Since flattening the spectrum does not change the topological equivalence class, $\aH$ is topologically nontrivial in the conventional sense, too. The class-AII Hamiltonian we obtain from the eigenvector matrix $U$ is then nontrivial as a class-AII ultra-localized insulator.

The TUI Hamiltonian obtained from this procedure has periodic boundary conditions with period $L_{x}$, $L_{y}$, $L_{z}$ in the $x$, $y$, and $z$ directions, respectively. We insert a flux $\Phi$ and write
\begin{equation}
  H(\Phi) = \sum_{\vn} E_{\vn} |\psi_{\vn}(\Phi)\rangle \langle \psi_{\vn}(\Phi)|.
\end{equation}
In order to bring about the hinge states, we set $E_{\vn} = E_x$ ($E_y$) if the localization center $\vn$ is within a localization length from the surface $x=0$ ($y=0$) and $E_{\vn} = 0$ otherwise. (We set $E_{\vn} = E_x$ if $\vn$ is within a localization length from both surfaces $x=0$ and $y=0$.) We then implement open boundary conditions in the $x$ and $y$ directions by setting all matrix elements $\langle \vr |H(\phi)| \vr' \rangle$ to zero for which $\vr$ and $\vr'$ are on opposite sides of the surfaces $x=0$ or $y=0$. This results in an open-boundary-condition  Hamiltonian $H_{\rm obc}(\Phi)$ defined for $0 \le x < L_x$ and $0 \le y < L_y$.
The spectrum of $H_{\rm obc}(\Phi)$ is shown as a function of $\Phi$ in Fig.~\ref{Fig:figure_hingestates}(b). It clearly displays Kramers-degenerate hinge modes along the hinges connecting the surfaces in $x$ and $y$ directions.
\end{widetext}

\end{document}